\documentclass[twocolumn,8pt]{IEEEtran}
\usepackage{amsfonts,amsmath,epsfig}
\usepackage{amssymb}
\usepackage{amsfonts}
\usepackage{multicol}
\usepackage{subfigure}
\usepackage{amsmath, amscd, dblfloatfix}
\usepackage{comment}
\usepackage{multirow}
\usepackage[belowskip=5pt, aboveskip=12pt]{caption}
\usepackage[normalem]{ulem} 

\begin{document}\sloppy

\def\x{{\mathbf x}}
\def\L{{\cal L}}

\title{Redundancy and Aging of Efficient Multidimensional MDS--Parity Protected \\ Distributed Storage
Systems}

\author{Suayb~S.~Arslan,~\IEEEmembership{Member,~IEEE}
\thanks{Suayb S. Arslan is with the Advanced Development Laboratory at Quantum Corporation, Irvine, CA, USA. E-mail: Suayb.Arslan@Quantum.com.}%
\thanks{Copyright $\copyright$ 2013 IEEE. Personal use of this material is permitted. However, permission to use this material for any other purpose must be obtained by sending request to pubs-permissions@ieee.org.}%
\thanks{Digital Object Identifier 10.1109/TDMR.2013.2293491}
}

\maketitle \setcounter{page}{1} \pagenumbering{arabic}

\begin{abstract}
The effect of redundancy on the aging of  an efficient Maximum Distance Separable (MDS)
parity--protected distributed storage system that consists of multidimensional arrays of storage units is explored. In light of the
 experimental evidences and survey data, this paper develops
generalized expressions for  the reliability of array storage systems based on more realistic time to failure distributions such as Weibull. For instance, a distributed disk array system is considered in which the array components
are disseminated across the network and are subject to independent failure rates. Based on such, generalized
closed form hazard rate expressions are derived. These expressions are extended to estimate the asymptotical reliability behavior of large scale storage networks equipped with MDS parity-based protection. Unlike previous studies, a generic hazard rate function is assumed,
a generic MDS code for parity generation is used, and an evaluation of the implications of
adjustable redundancy level for an efficient distributed storage system is presented. Results of this study are applicable to any erasure
correction code as long as it is accompanied with a suitable structure and an appropriate
encoding/decoding algorithm such that the MDS property is
maintained.
\end{abstract}

\markboth{This paper has been accepted for publication in IEEE Transactions on Device and Materials Reliability,  Nov. 2013 }%
 {Shell
\MakeLowercase{\textit{Arslan at al.}}:QTM Advance Channel Group}

\section{Introduction}
\label{sec:intro}

One of the
well known problems associated with parity-based redundant array of
inexpensive disk (RAID) systems \cite{Patterson} is their vulnerability against
multiple disk failures, mostly after which a restore mechanism is initiated
and subsequent read errors inevitably occur. Similar trends can be observed in arrays of solid state drives (known as RAIS) for mass storage applications \cite{RAIS}. The statistical likelihood of multiple drive
failures has never been a significant issue in the past. Over the years however, with the advanced technology, drives  of few terabyte capacities are now put on sale.  The scale of storage systems continues to grow to
store peta-bytes of data and the likelihood of multiple drive failures become
a reality. This led to the development of error checking and validation routines to maintain the data integrity. Conventional approach for data retention was to address the big data protection shortcomings of RAID by replication, a technique of making additional copies of data to
avoid unrecoverable errors and lost data. Organizations also used replication schemes to help with failure scenarios, such as
location specific failures, power outages, bandwidth unavailability, and so forth. However, as the size of the stored data scales up, the number of copies of the data required for robust protection grows. This increases the amount of inefficiency by adding extra cost to the overall system. Since replication leads to extremely inefficient use of system resources, parity-based protection using error correcting codes is  more popular.

Drive failures can be regarded as arrivals of
a renewal process at a certain rate.
The drive failure rate,  using a homogenous Poisson process, is the reciprocal
of the mean time to failure (MTTF) numbers reported by the drive
manufactures \cite{IDEMA}. One of the earliest studies of the reliability
analysis for disk array systems considered various RAID hierarchies
and hot spots \cite{RAID1}. In a number of successive works, stripping is used as to provide
cost-effective I/O systems \cite{Patterson}, \cite{Salem} for seemless and reliable access to the user data. Most of
the previous research modelings were based on single or double--parity schemes such as RAID 5 or RAID 6 in which maximum distance separable (MDS) codes are used for storage efficiency. MDS codes have the nice property that for a given array and parity size they allow maximum amount of recovery  \cite{Blahut}. However, these studies mostly
assume constant failure rate of small size and cost--effective disk
components. Unfortunately, these set of assumptions are shown to be unrealistic \cite{Gibson}, \cite{1000}. In fact, an interesting observation is that the failure rates are rarely constant \cite{1000}, \cite{Shah}.

There have been efforts in industry as well as in academia
for accurately predicting the reliability of large scale storage
systems in terms of
mean lifetime to failure rates. For example, an accurate yet complicated model is developed to include catastrophic failures and usage dependent data
corruptions in \cite{RAID1REL}. The authors specifically pointed out that component failure rates have little, if not any, to share with the failure rate of the whole storage system. The times between successive system failures are reported to be relatively larger than what conventional models suggest, even though each component disk hazard rate is increasing \cite{Ascher}. Disk scrubbing is introduced and used in \cite{scrub} as a remedy for latent defects that are usually independent of the size, use and the operation of disks. The latter study also uses homogenous Poisson model for reliability estimations.

It is clear that excessive failures (failures beyond the correction capability of the system) in any storage system are of
particular interest because they may cause both unavailability and
permanent data loss. On the other hand, the trend in the market is to grow
the scale of distributed storage arrays  in which the capacity as well as the
reliability of each storage component almost double every year. Therefore, a true and accurate failure modeling shall be of great significance from a system design standpoint. For example, a generic hazard rate function $\lambda(x)$ and an associated non-homogenous Poisson model might be a better fit for predicting the real life disk failure trends. However, as more real life scenarios are incorporated with these improved mathematical methods for accuracy, they inexorably become complex. From a customer's perspective, short-hand closed form expressions for predicting the system failure rates might be more useful for delivering performance figures about the  system reliability.

In this study, an efficient MDS-parity based
distributed disk array  system is considered using general failure processes.
One of the contributions of this paper is a set of useful closed form expressions, derived by considering the whole lifespan of component drives based on
the recent survey data on disk failures and time to failure
probability distributions \cite{1000}, i.e., without assuming constant component hazard rates. Some asymptotical results (the array size tends to infinity) shed light for the limiting behavior of RAID type systems. Those results might particularly be important for predicting what is achievable using MDS codes, as the scale of the coded storage systems grow for the management/maintenance of the so called ``Big data".
The paper also investigates the relationship between the aging and the redundancy used for data protection.  Here,  the efficiency of the distributed storage system comes rather from the efficient allocation strategy such that independent drive failure
assumption is roughly correct for each component of the array, which are shared by different storage network nodes. It is further shown that the multidimensional array storage is offering a good tradeoff between complexity and performance which may otherwise be obtained by a large array of one dimensional RAID type systems at the expense of increased cost and complexity. Although the main objective of the paper is focused on the mean time to first failure, the expressions can be extended to mean time between failures and mean time to data loss performance metrics. However, derived expressions might either not be in simple form or expressible in a closed form for an arbitrary hazard rate function $\lambda(x)$ and a repair function $\mu(x)$.

The remainder of this paper is organized as follows: In Section II,
a brief introduction is given about the reliability theory basics as well as the drive failure statistics in real world. Moreover,  the storage system details is summarized along with the assumptions used in this work.
In Section III, main results of the paper are given based on arbitrary hazard rates using multidimensional arrays. This section starts with considering 1-D arrays and then generalizes the results for multidimensional arrays.
Some of the numerical results and relevant examples are given in Section IV. Finally, a brief summary
and conclusions follow in Section V. The proofs are included in appendices A, B and C in order to highlight the main contributions of the paper.

\section{Background and System diagram}

\subsection{Reliability Theory}

When a brand new product is put into service, it performs functional operations satisfactorily  for a  period of
time, called \emph{useful time} period, before eventually a failure occurs and the device is no longer able to
respond to user requests. The observed time to failure ($TTF$) is a
continuous random variable with a probability density function
$f_{TTF}(x)$, representing the lifetime of the product until the first failure. The
 failure probability of the device can be found using the
cumulative distribution function (CDF) of $TTF$  as follows,
\begin{eqnarray}
F_{TTF}(x) = Pr\{TTF \leq x\} = \int_{0}^{x} f_{TTF}(y) dy, \ \ x
> 0
\end{eqnarray}

We can think of $F_{TTF}(x)$ as an \emph{unreliability} measure between
time 0 and $x$. The reliability function $S(x)$
is therefore defined by,
\begin{align}
S(x) \triangleq 1 - F_{TTF}(x)
= \int_{x}^{\infty} f_{TTF}(y) dy \label{Rel}
\end{align}

In other words, reliability is the probability of having no failures before
time $x$ and is related to CDF of $TFF$ through  Eqn. (\ref{Rel}).
Note that Eqn. (\ref{Rel}) implies that we have
$f_{TTF}(x)=-dS(x)/dx$. It may not be possible to estimate the
distribution function of $TFF$ directly from the available physical
information. A useful function in clarifying the relationship
between physical modes of failure and the probability distribution
of $TFF$ is known as the \emph{hazard rate} function or
\emph{failure rate} function, denoted as $h_{TTF}(x)$. This function
 is defined to be of the form
\begin{eqnarray}
h_{TTF}(x) \triangleq \frac{f_{TTF}(x)}{S(x)}  = -\frac{dS(x)}{S(x)dx}  \label{ODE}
\end{eqnarray}

\begin{figure}[b!]
\centering
    \includegraphics[width=80mm, height=56mm]{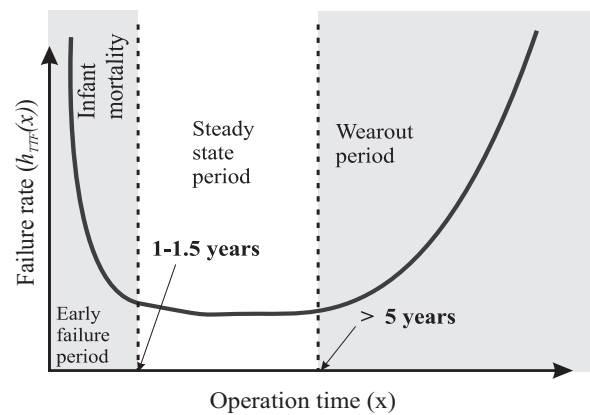}
    \captionof{figure}{\footnotesize Hazard rate pattern for hard disk drives as a function of operation time \cite{Yang}}
    \label{bathtub}
\end{figure}

\begin{figure*}[t!]
\centering
    \includegraphics[width=152mm, height=65mm]{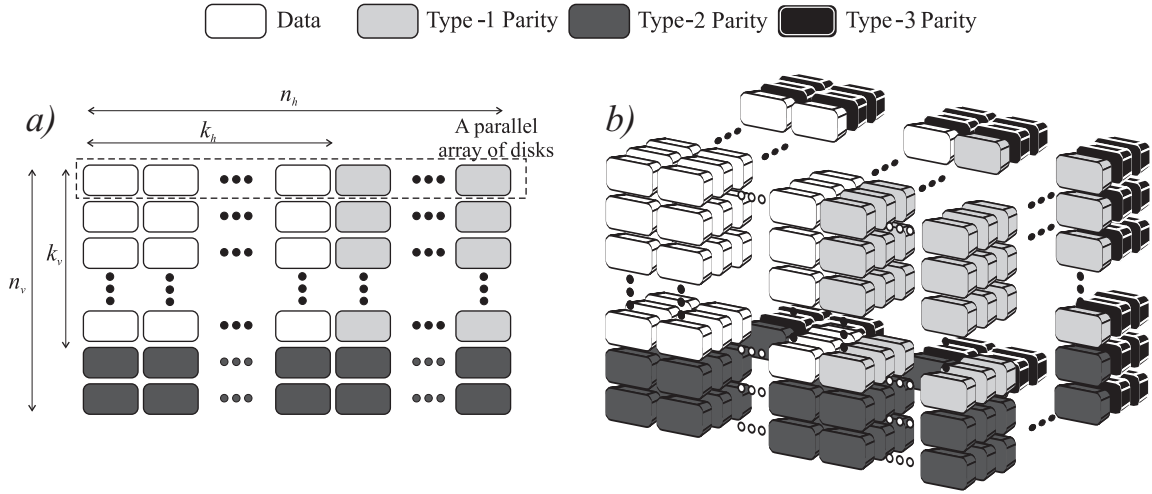}
    \captionof{figure}{\footnotesize Assume that the storage units are disks. a) Parallel series of disk arrays that  that makes up one disk matrix to be distributed over the network nodes. Each block represents a disk belonging to one of the parity types \cite{Yang}. b) A set of these disk matrices are used for storing large scale data using MDS encoding along each dimension.}
    \label{Matrix}
\end{figure*}

Solution of the first order ordinary differential equation
(\ref{ODE}) yields the relationship $h_{TTF}(x) =-d(\ln(S(x)))/dx$ with the initial condition $S(0)=1$. Note that  knowing the hazard rate is
equivalent to knowing the distribution. Mean time to failure (MTTF) is defined to be the expected value of the random
variable $TTF$ and is given by
\begin{align}
MTTF &\triangleq& \mathbb{E}[TTF]
= \int_{0}^{\infty} S(x) dx \Leftrightarrow \lim_{x \rightarrow \infty} x S(x) = 0 \label{IFF}
\end{align}
where $\mathbb{E}[.]$ is the expectation operator. Note that Eqn. (\ref{IFF}) is true for distributions whose mean exists.
 For the rest of our discussion, the subscript $TTF$ is dropped for notation simplicity and throughout the text, $h_{TTF}(x)$ is alternatively denoted by $\lambda(x)$.

Annualized failure rate (AFR) is frequently used to estimate the failure probability of a
device or a component after a full-time year use. In the
conventional approach, time between failures are assumed to be independent and
exponentially distributed with a constant rate $\lambda$. Therefore AFR is given by $AFR = 1 - S(x) = 1 - e^{-\lambda x}$,
where $\lambda = 1/MTTF$ and $x$ is the running
time index in hours. MTTF is reported in hours and since there are
8760 hours in a year, $AFR = 1 - e^{-8760/MTTF}$. Typical numbers reported in disk vendor's specifications are
$MTTF \approx $ 1 million hours or 1.5 million hours. Since $8760/MTTF \ll 1$,
 $AFR \approx  8760 / MTTF$. As mentioned before, such rough calculations and assumptions  may not be representing how the
drives behave in real world \cite{1000}. Clearly, one needs general but yet adequately simpler expressions to predict the lifetime trends of such systems, particularly for large scale storage applications.

\subsection{Drive failures in real world}

It is shown in various research articles
 that average replacement rate of component disks is around twenty times much
greater than are the theoretically predicted MTTF values i.e., predicted MTTF values are observed to be an
underestimator for this scenario \cite{1000}, \cite{Yang}.   It is
demonstrated that disk failure rates show a ``bathtub" curve as
shown in Fig. \ref{bathtub}. Additionally, contrary to conventional homogenous stochastic models, hard disk
replacement rates do not enter into steady state. After few years of
use, drives (majority of which are disks) are observed to enter into wear-out period in which the
failure rates steadily increase over time. Time between
failures are shown to give much better fit with Weibull or gamma
distributions instead of widely used exponential distribution \cite{1000}. There is a considerable
amount of evidence that disk failures that are placed in the same batch show significant correlations, which is
hard to quantify in a number of applications \cite{Pinheiro}.

\subsection{Efficient storage system summary}

A series of parallel array of storage units (such as disk drives) is shown in Fig. \ref{Matrix} a). Drives are assumed
to be manufactured identically and share the same
 failure/hazard rate function $\lambda(x)$. The $k_h \times k_v$ data matrix is encoded using two different block MDS codes in order to create a 2-D array. Horizontally, the parity information type--1 is computed using a $(n_h,k_h,t_h+1)$ MDS
code which can correct up to  $t_h$ erasures per block. It is due to the MDS property that it is the maximum number of erasures that a $(n_h,k_h)$ block code can correct. Then, the computed
parities are allocated to different disk units and occupy a fraction
of the storage space of the disk array to  protect the system
against various types of system and disk failures. In addition to horizontal encoding, the parity information type--2
is computed using another $(n_v,k_v,t_v+1)$ MDS code which can correct up
to  $t_v$ erasures per vertical block. This encoding procedure can be performed repeatedly to protect larger dimensional data sets. The order of encoding does not matter as long as the code is a linear block code. A generalization of such an encoding scheme for three dimensional data is depicted in  Fig. \ref{Matrix} b). Finally, encoded data units are allocated into the network storage nodes according to a genuine allocation policy that
will keep
\begin{itemize}
\item[(\emph{i})] the read and write process simple,
\item[(\emph{ii})]  the number of storage nodes needed to be accessed for the reconstruction of user/parity data
minimum at a reasonable time complexity
\item[(\emph{iii})] drives in the horizontal or
vertical  arrays (for 2-D array case) not shared by the same node of the
distributed storage system.
\end{itemize}

These set of assumptions also help us make independent failure  assumptions between the component drives of an array while in the mean time facilitate the rest of our analysis.

\section{Disk arrays with individual independent arbitrary hazard rates}

In the rest of our discussions, an allocation policy and a generic MDS code  are  assumed such that conditions \emph{(i)}, \emph{(ii)} and \emph{(iii)} are satisfied. Therefore, the rest of the discussion is based on the independent failure statistics assumption between respective drives of the storage array. A series of parallel arrays of disks is considered in which each array contains $n$
disks or drives to store the encoded user data information. Let us use a
common notation $(n,k,t+1)$ where $t = n - k$ for the MDS code in order to make it
general and applicable to each and every dimension to which erasure coding is applied.

\subsection{A Horizontal system and componentwise reliability}

Let us consider a 1-D array of storage units. Note that the results of this subsection can be applied to other arrays of different dimensions.  This subsection starts with stating our main theorem  below that bridges the relationship between redundancy and
aging of MDS parity-based arrays.

\textbf{Theorem 1:} \emph{Hazard rate per data component of a
horizontal system (consisting of $n$ independent components $k$ of
which are data, each component having an arbitrary but the same
failure rate of $\lambda(x)$), coded with a generic $(n,k,t+1)$
block MDS code with rate $r=k/n$ ($t=n-k$ parity units) is given by}
\begin{eqnarray}
\mu_c(x,n,r) &=&
\frac{\lambda(x)}{r}\left(1 -\frac{\psi_{t-1}(n-1,\lambda(x))}{
\psi_{t}(n,\lambda(x))}\right) \label{Thm1Core}
\end{eqnarray}
\emph{where}
\begin{eqnarray}
\psi_t(z,n, \lambda(x)) &\triangleq& \sum_{i=0}^t \binom{n}{i}
\binom{i}{z} (1-R(x))^i R(x)^{n-i}, \nonumber
 \end{eqnarray}
\emph{$R(x) = e^{-\int_0^x\lambda(y)dy}$ is the reliability
of constituent components with hazard rate $\lambda(x)$ and $\psi_{t}(n, \lambda(x)) \triangleq \psi_{t}(0, n,
 \lambda(x))$ is the cumulative distribution function of the binomial distribution. Furthermore, the following inequality is satisfied for $0 \leq t \leq n-1$}
\begin{eqnarray}
 \max\left\{0,\frac{1-R(x)/r}{1-R(x)} \right\}  \leq  \frac{\mu_c(x,n,r)}{\lambda(x)} \label{FurthermoreThm1}
\end{eqnarray}

 \textbf{\emph{Proof}:}   See Appendix A.

Two cases shed some interesting light to this relationship. Consider the case with $t=0$ and $r=1$,
i.e., no redundancy. In this case, since
$\psi_{t-1}(n-1,\lambda(x))=0$ we have $\mu_c(x,n,1) = \lambda(x)$ as expected. In
other words, the hazard rate of the horizontal system per component is
the same as the hazard rate of the constituent components when there
is no redundancy. On the other extreme, we could have $t=
n-1$ and  $r=1/n$  using replication. In this case, the hazard rate per data component is given by
the following \emph{corollary}.

\textbf{Corollary 2:} \emph{Using a $(n,1,n)$ block MDS code, known
as repetition code, the hazard rate per data component is given by}
\begin{eqnarray}
\mu_c(x,n, 1/n) = \frac{n\lambda(x)R(x) (1-R(x))^{n-1}}{1-(1-R(x))^n}
\end{eqnarray}
\emph{where  $R(x) = e^{-\int_0^x\lambda(y)dy}$ is the reliability
of constituent components with hazard rate $\lambda(x)$.}

\textbf{\emph{Proof}:} Let us set $t=n-1$, we have
\begin{eqnarray}
\psi_{n-1}(n, \lambda(x)) &=& \psi_{n}(n,\lambda(x)) - (1-R(x))^n \nonumber \\
&=& 1 - (1-R(x))^n  \label{cor_eqn_1}\\
\psi_{n-2}(n-1, \lambda(x)) &=& \psi_{n-1}(n-1,\lambda(x)) - (1-R(x))^{n-1} \nonumber \\
&=& 1 - (1-R(x))^{n-1} \label{cor_eqn_2}.
\end{eqnarray}

The result will follow through some algebraic manipulations by plugging Eqns. (\ref{cor_eqn_1}) and (\ref{cor_eqn_2}) into Eqn.
(\ref{Thm1Core}). \hfill $\blacksquare$

A well known binary linear block MDS code is the parity code
in which there is only one parity symbol i.e., $t=1$ and $r=(n-1)/n$. Following
corollary characterizes the hazard rate of 1-D array using parity
coding.

\textbf{Corollary 3:} \emph{Using a $(n,n-1,2)$ binary block MDS code,
known as parity code, the hazard rate per data component of a
horizontal block is given by}
\begin{eqnarray}
\mu_{c}(x,n,1-1/n) = \frac{\lambda(x) n(1-R(x))}{n(1-R(x))+R(x)}
\end{eqnarray}
\emph{where  $R(x) = e^{-\int_0^x\lambda(y)dy}$ is the reliability
of constituent components with failure rate $\lambda(x)$.}

\textbf{\emph{Proof}:} We recognize that for $t=1$ and,
\begin{eqnarray}
\psi_{0}(n-1, \lambda(x)) &=&R(x)^{n-1} \label{cor1} \\
\psi_{1}(n, \lambda(x)) &=& R(x)^{n} + n (1-R(x))R(x)^{n-1}
\label{cor2}
\end{eqnarray}

By plugging Eqns. (\ref{cor1}) and (\ref{cor2}) into Eqn.
(\ref{Thm1Core}), with $r=1-1/n$, we have
\begin{align}
\mu_{c}(x,n,1-1/n) =& \frac{\lambda(x)}{r} \left(1 - \frac{1}{R(x) +
n(1-R(x))}\right)
\\ =& \frac{\lambda(x)n}{n-1} \left( \frac{(n-1)(1-R(x))}{R(x)
+ n(1-R(x))} \right)
\\ =& \frac{ \lambda(x) n(1-R(x))}{n(1-R(x))+R(x)}
\end{align}
as desired. \hfill $\blacksquare$

In Theorem 1, if $R(x) \geq r$, the lower bound becomes zero, whereas if $R(x) < r$, the lower bound takes on a non-zero value. Let us define a system to be \emph{componentwise reliable} (CR) if the hazard rate per drive component is zero or close to zero although the hazard rate of the whole system might be non-zero. Therefore, an interesting question is whether the lower bound of Theorem 1 is achievable for any real value of $R(x)$ and $r$ as $n$ grows large. This question will be explored next.

\begin{figure*}[t!]
\centering
\includegraphics[width=88mm, height=59mm]{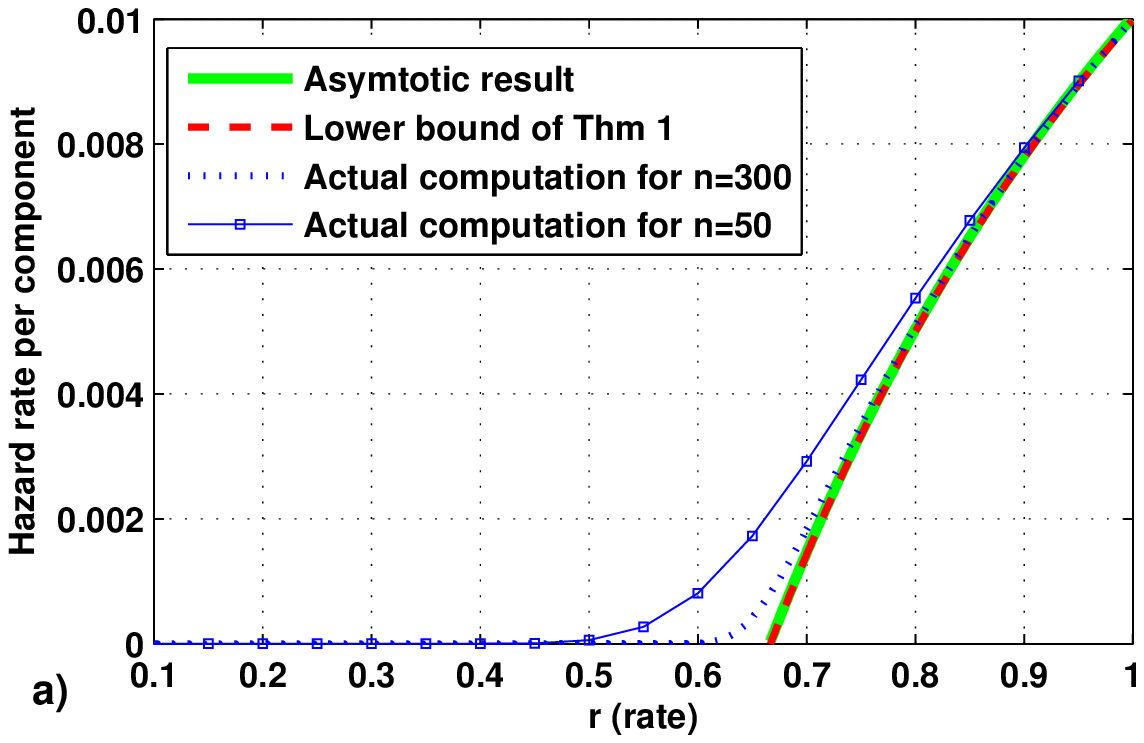}
\includegraphics[width=88mm, height=59mm]{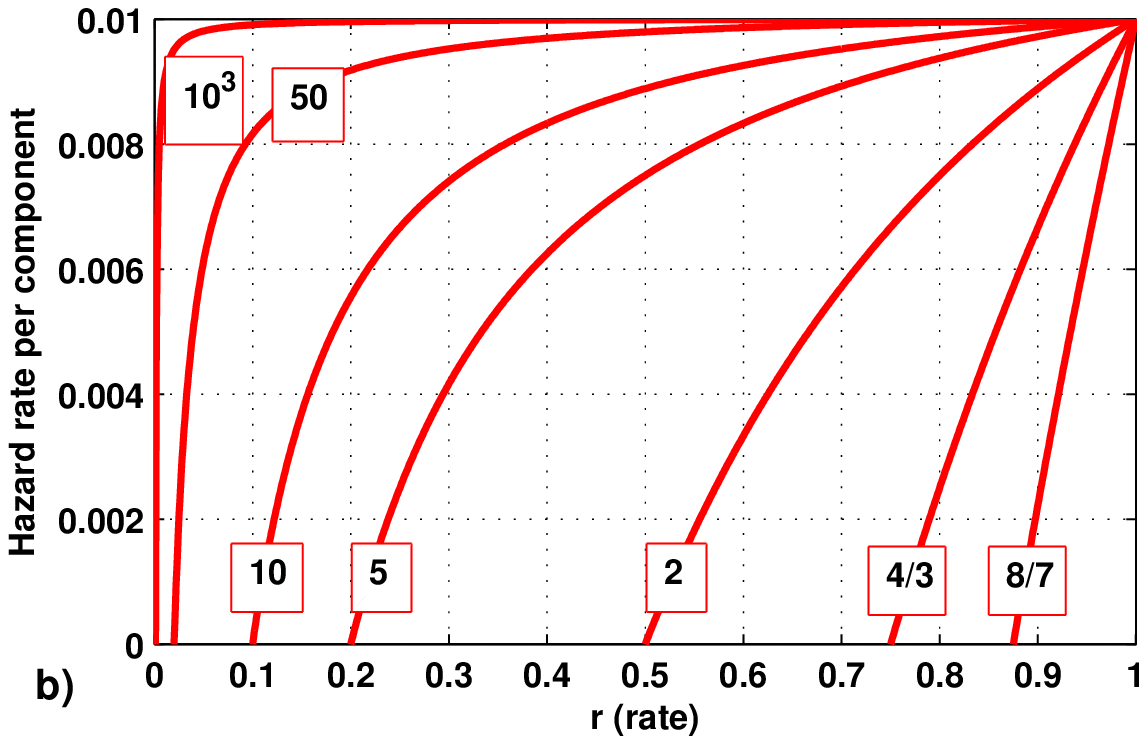}
\captionof{figure}{\footnotesize a) Asymptotic per component hazard rate when $R(a) = 1/q$ where $q = 3/2$ using an MDS code.  b) Asymptotically achievable bound as a function of rate $r$ and $q$. As can be seen as $q$ gets larger, the hazard rate per component tends to $\lambda(x)$ and become less dependent on rate. $\lambda(a)=0.01$ is assumed.}
\label{Aym5}
\end{figure*}

\subsection{Asymptotical hazard rate expressions}

Regardless of the reliability distribution model
used, as $x \rightarrow \infty$, the reliability of constituent
components, $R(x)$ tends to zero. Therefore, one of the information
Theorem 1 conveys is that for a fixed block length of $n$, if we let
$R(x) \rightarrow 0$, we have the following lower bound,
\begin{eqnarray}
\lim_{x \rightarrow \infty} \mu_c(x,n,r) \geq  \lambda(x).
\end{eqnarray}

Therefore, one might expect gains due to erasure coding in the infant mortality and useful time period, but not much in long term wear-out periods for 1-D arrays.
As the number of component disks increases with the
growing need for big data storage, the corresponding system reliability
might be going down. It is of interest therefore to look at the
asymptotic behavior $n
\rightarrow \infty$ of these reliability expressions at different times $x$.

Let us start with evaluating the limiting behavior $n
\rightarrow \infty$ for a finite non-zero value of $a$ such that $R(a) = 1/q$ for $q=2, 3, \dots$. In other words, for a given $\lambda(x)$, find $a$ such that $\int_{0}^a \lambda(y) dy = \ln q$. For this special case, we have the following asymptotic result that shows the lower bound of Theorem 1 is achievable.

\textbf{Theorem 4:} \emph{Asymptotic hazard rate per data component
of a horizontal system (as $n \rightarrow \infty$, each component
having an arbitrary but the same failure rate of $\lambda(x)$), coded
with a generic $(n,k,t+1)$ MDS block code with a fixed rate $r=k/n$ is given by
\begin{eqnarray}
\mu_c(a, n ,r) = \begin{cases}
\frac{\lambda(a)(qr-1)}{r(q-1)} &\mbox{if } r \geq R(a) = \frac{1}{q}  \label{Thm5Core1}  \\
0  &\mbox{Otherwise}
\end{cases}
\end{eqnarray}
where $a$ satisfies the relationship $\int_{0}^a \lambda(y) dy = \ln q$ for a given positive integer $q$.
}

\textbf{\emph{Proof}:} See Appendix B. The proof also conjectures that this theorem can be extended to any $q \in \mathbb{R}, q > 1$.

For $r \geq 1/q$, let us divide both the numerator and denumerator by $qr$ and replace $q$ with $1/R(a)$. We will have $\mu_c(a, n ,r) = \lambda(a) \left(\frac{1-R(a)/r}{1-R(a)}\right)$. Yet, this is the lower bound predicted by Theorem 1 evaluated at point $x=a$. Therefore, Theorem 4 proves that the lower bound of Theorem 1 is achievable for a countably infinite number of values of $R(x)$. We also conjecture that the lower bound of Theorem 1 is achievable for any value of $R(x)$ i.e., for any $q \in \mathbb{R}, q > 1$. Let us provide an example to support this conjecture by setting $q=3/2$, a non-integer value, and $\lambda(arg \min_{a} \left\{ | \int_{0}^a \lambda(y) dy - \ln q | \right\}) = 0.01$  is fixed for simplicity. We plot the asymptotic result, the lower bound due to  Theorem 1  as well as the actual computation in Fig. \ref{Aym5} a). As can be seen asymptotic result of  Theorem 4 achieves the lower bound of Theorem 1 for $r  \geq 2/3$, below which we have a CR system if $n$ is very large. However, if $n=50$ or $n=300$ we observe some performance loss and in order to obtain a CR system we must have $r \leq 0.12$ and $r\leq 0.46$, respectively.

So far, we have assumed that the size of the array $n$ is increased for a given fixed value of $R(x)$. If $R(x) \rightarrow 0$ i.e., $q \rightarrow \infty$, we can see that the hazard rate per component of an MDS-protected array will converge to $\lambda(x)$. This is shown for a fixed value $\lambda = 0.01$ in Fig. \ref{Aym5} b). Yet, a general practice should be adaptively increasing the size $n$ as $R(x)$ tends to zero i.e., as the reliability of components go down with time. The following theorem characterizes this scenario with the assumptions of an adaptive system: $\lim_{\substack{ n \rightarrow \infty \\ x \rightarrow \infty}}
nR(x) < \infty$ and $\lim_{\substack{ n \rightarrow \infty \\ x \rightarrow 0}}
n(1-R(x)) < \infty$ and shows that a CR system is possible for large scale storage even if the component drives are in their wear-out period.

\textbf{Theorem 5:} \emph{if $\lim_{\substack{ n \rightarrow \infty \\ x \rightarrow \infty}}
nR(x) < \infty$ and $\lim_{\substack{ n \rightarrow \infty \\ x \rightarrow 0}}
n(1-R(x)) < \infty$ , asymptotic hazard rate per data component
of a horizontal system (as $n \rightarrow \infty$, each component
having an arbitrary but the same failure rate of $\lambda(x)$), coded
with a generic $(n,k,t+1)$ MDS block code with a fixed rate $r =
k/n$ is given by}
\begin{eqnarray}
\lim_{\substack{ n \rightarrow \infty \\ x \rightarrow a}}
\mu_c(x,n,r)  =
\frac{\lambda(x)C(x,r,a)}{n}
\end{eqnarray}
\emph{where}
\begin{eqnarray}
C(x,r,a) = \begin{cases}
1/r &\mbox{if } a = \infty \\
\frac{1-R(x)}{(R(x)-r)(2R(x)- r - 1)} &\mbox{if } a = 0
\end{cases} \nonumber
\end{eqnarray}
\textbf{\emph{Proof}:}   See Appendix C.

The amount of
redundancy has a positive effect on the component hazard rate
$\lambda(x)$ for this particular scenario. For a fixed $n$, the hazard rate for
overall disk array was found to be scaling with $rn\lambda(x)$ if $n \rightarrow \infty$ first, then $R(x) \rightarrow 0$. On the other hand if $R(x) \rightarrow 0$ and $n \rightarrow \infty$ at the same time such that their product stays constant, this hazard rate converges to $\lambda(x)$, i.e., $k=rn$ times
less than that of the fixed $n$ case. Large block length improves the reliability performance at the expense of increased complexity. Note also that although the per component hazard rates might be tending to zero, the overall array hazard rate is nonzero.

The results of this subsection establishes an important relationship between the concept of CR and the rate $r$ of the MDS code used. However, we assumed that the rate $r$ is fixed through the whole lifespan of the storage system. Thus, it is easy to see that at some point in time $x^{\prime}$ we will have $r > R(x^{\prime})$ and $\mu_c(x^{\prime}, n ,r) \not= 0 $ even if $n \rightarrow \infty$. Thus in order to obtain a CR system at all times, MDS codes with time varying rate $r(x)$ might be quite useful. In fact, there are asymptotically MDS codes called fountain codes that can be a perfect fit for this particular scenario \cite{MacKay}. Using such codes, rate can be adjusted on the fly such that $r(x) \leq R(x)$ is satisfied for all $x$ if the condition of CR is strictly imposed on the design throughout the lifespan of the storage system.

\subsection{Multidimensional disk arrays}

Although the potential for a CR system is shown using large 1-D MDS-protected arrays, the implementation details and real life conditions make it impractical to achieve idealized performance benefits. Therefore, different directions must be taken for practical means such as multidimensional arrays using MDS codes. This is one of the natural ways to construct long blocks of many drives that can help us realize the asymptotical results derived in the previous subsection.

Previous section considered replaceable drive components in a 1-D horizontal
structure and posed the question for any type of MDS code of rate
$r_h$. Let us assume that we have a series of such parallel blocks
of drives of hazard rate $k_h\mu_c(x,n_h,r_h)$ as shown in Fig. \ref{Matrix} a),
generated by another MDS code of rate $r_v$. In other words, we have
$n_v$ 1-D arrays of size $n_h$  drives each, such that if more than $n_v(1-r_v)$
blocks fail, it will lead to the whole system failure. This is due to the MDS property of the erasure correction coding. Furthermore, we assume inter-block failure independence and if a horizontal block fails, all the constituent disks are assumed to be failed. For a general case, this type of decoding procedure corresponds to the failure of T-D disk array, if at least one (T-1)-D disk array fails. More complicated decoding procedures can be employed for better performance at the expense of increased implementation complexity.

Let
$\mu_{c}(x,n_hn_v,r_hr_v)$ be the hazard rate per data component of the 2-D disk array. Using the result of
Theorem 1, we shall obtain
\begin{align}
& \mu_{c}(x,n_hn_v,r_hr_v) =  \frac{\mu_c(x,n_h,r_h)}{k_hr_v} \nonumber \\
& \ \ \ \ \ \ \ \ \ \ \ \ \ \times \left(1 -\frac{\psi_{(1-r_v)n_v-1}(n_v-1,k_h\mu_c(x,n_h,r_h))}{
\psi_{(1-r_v)n_v}(n_v,k_h\mu_c(x,n_h,r_h))}\right)
\end{align}

The result follows
from Theorem 1 by replacing $\lambda(x)$ with $k_h\mu_c(x,n_h,r_h)$ and $n$ with $n_v$. Finally,  the system failure rate is divided by the total number of data disks in a horizontal array. This expression is indeed a special case of the following more general result on T-D disk array system encoded with a set of MDS codes with parameters
\{$(n_1,r_1)$,$(n_2,r_2)$,$\dots$,$(n_T,r_T)$\}. Note that 3-D case is shown in Fig. \ref{Matrix} b) and larger dimensional generalizations are possible yet are hard to visualize.

\textbf{Theorem 6:} \emph{For T-D MDS-protected system of drives or disks, we have
\begin{align}
 & \mu_{c}(x,n_{1,T},r_{1,T})=   \frac{\mu_{c}(x,n_{1,T-1},r_{1,T-1})}{r_{1,T}n_{1,T-1}} \\ \times
& \left(1 -\frac{\psi_{(1-r_T)n_T-1}(n_T-1,k_{1,T-1}\mu_{c}(x,n_{1,T-1},r_{1,T-1}))}{
\psi_{(1-r_T)n_T}(n_T,k_{1,T-1}\mu_{c}(x,n_{1,T-1},r_{1,T-1}))}\right)
\end{align}
where $n_{1,s} \triangleq \prod_{i=1}^s n_i$, $r_{1,s} \triangleq \prod_{i=1}^s r_i$ and $k_{1,s} = r_{1,s}n_{1,s}$.}

\textbf{\emph{Proof}:} (sketch) Proof follows from Theorem 1 by replacing $\lambda(x)$ with $k_{1,T-1}\mu_{c}(x,n_{1,T-1},r_{1,T-1})$ in which $\mu_{c}(x,n_{1,T-1},r_{1,T-1})$ is the hazard rate per component  for (T-1)-D parity protected system and $k_{1,T-1}$ is the number of data component disks. In this case, the result of Theorem 1 can be  applied to compute the hazard rate of a disk hyperplane of dimension T-1. In order to find the hazard rate per  component, we divide the overall hazard rate function by the number of components $k_{1,T-1}$. We eventually obtain the result using the fact $r_{1,T-1}r_T = r_{1,T}$. \hfill $\blacksquare$

\section{Examples and Numerical Results}

In this section,  results will be provided for some of the special cases for finite block lengths so that a comparison can be made with asymptotical results. Reliability of multidimensional arrays will be compared in terms of component as well as array level hazard rates.

\subsubsection{Constant hazard rate components with $r=1/n$}

Consider a parallel block and the non-aging components with a
constant and identical rate i.e., $\lambda(x) = \lambda$. Using Corollary 2 and $R(x) = e^{-\lambda x}$ for constant
rate $\lambda(x) = \lambda$, as derived in
\cite{Rel}, we have
\begin{eqnarray}
\mu_c(x, n, r) = \frac{n\lambda e^{-\lambda x} (1-e^{-\lambda
x})^{n-1}}{1-(1-e^{-\lambda x})^{n}} \label{ref8}
\end{eqnarray}

Note that this constant failure rate assumption was originally used by manufactures to predict the failure trends.
 Equation (\ref{ref8}) can be approximated as $\mu_c(x, n, r)  \approx n
\lambda^n x^{n-1}$ when $x \ll 1/ \lambda$. As $x \rightarrow \infty$, the hazard
rate converges to $\lambda$, achieving the lower bound of Theorem 1. In its early
use, the system failure rate grows as a power function of age which
is known as the Weibull law. This means that using more redundancy within the block triggers
aging although the constituent parts are non-aging components.

\subsubsection{Non-constant hazard rate components using arbitrary $r$}

Let us consider a general form of $\lambda(x)$ that is in bathtub shape using a composite distribution model\footnote{ A Weibull hazard function is used to model three different periods of failure processes by appropriately choosing the parameters of the distribution.} given by
\begin{align}
\lambda(x) = \begin{cases} \frac{\beta_1t^{\beta_1-1}}{\theta_1^{\beta}} &\mbox{if } 0 < t \leq t_1 \\
\frac{\beta_2t^{\beta_2-1}}{\theta_2^{\beta_2}} & \mbox{if } t_1 < t \leq t_2 \\
\frac{\beta_3t^{\beta_3-1}}{\theta_3^{\beta_3}} & \mbox{if } t > t_2 \end{cases}  \label{EqnMix}
\end{align}
with the corresponding reliability function,
\begin{align}
R(x) = \begin{cases} e^{-\left(\frac{x}{\theta_1}\right)^{\beta_1-1}} &\mbox{if } 0 < t \leq t_1 \\
e^{-\left(\frac{x-t_1}{\theta_2}\right)^{\beta_2-1} - \left(\frac{t_1}{\theta_1}\right)^{\beta_1-1}} & \mbox{if } t_1 < t \leq t_2 \\
e^{-\left(\frac{x-t_2}{\theta_3}\right)^{\beta_3-1} - \left(\frac{t_2-t_1}{\theta_2}\right)^{\beta_2-1} - \left(\frac{t_1}{\theta_1}\right)^{\beta_1-1}} & \mbox{if } t > t_2 \end{cases}  \label{EqnMix}
\end{align}

\begin{figure}[t!]
\centering
    \includegraphics[width=88mm, height=60mm]{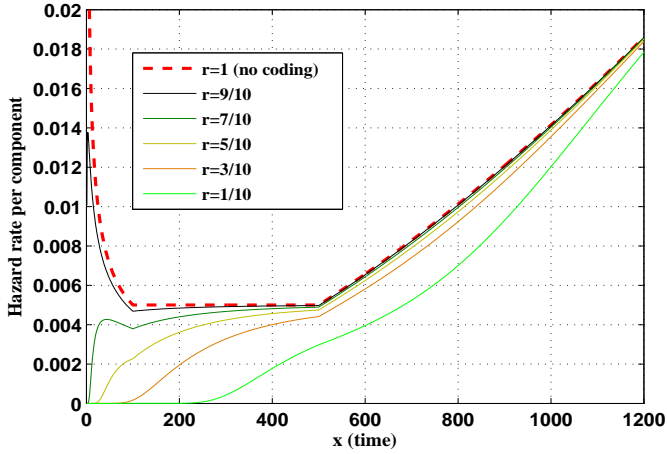}
    \captionof{figure}{\footnotesize Hazard rate function per component for 1-D disk array system as a function of time with and without coding. $\lambda(x)$ is assumed to be a simple bathtub curve obtained by using a composite distribution based on Weibull models. Let us assume $n=100$ disks per array. }
    \label{ComputeHazard}
\end{figure}

\begin{figure}[t!]
\centering
    \includegraphics[width=88mm, height=60mm]{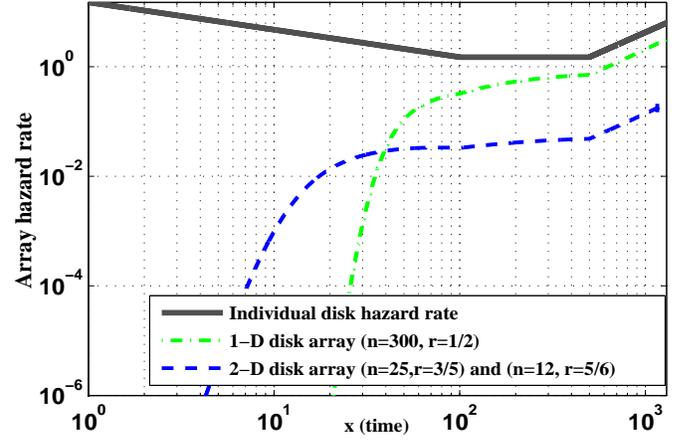}
    \captionof{figure}{\footnotesize Array hazard rate function using different MDS codes and structures. $\lambda(x)$ is assumed to be a simple bathtub curve obtained by using a composite distribution based on Weibull assumption. Number of disks in each system is 300 and the total rate of each MDS code is 1/2.}
    \label{ComputeHazard2}
\end{figure}

For useful life period (random failure process) between time $t_1$ and $t_2$, let us set $\beta_2 = 1$ and $\theta_2 = 200$. For the early life (infancy), we use $\beta_1 = 0.5$ and $\theta = 100$ to model the decreasing failure rates. In order to model the wear-out period, let us set $\beta_3 = 2.5$ and $\theta_3 = 500$.  An example is shown in Fig. \ref{ComputeHazard} for an array of disks  ($n=100$) using MDS codes with different rates.
As predicted by asymptotical expressions, for $x \rightarrow 0$ we have $\mu_c(x,100,r) \rightarrow 0$ and  for $x \rightarrow \infty$ we have $\mu_c(x,100,r) \rightarrow \lambda(x)$. This figure also suggests that there is a key number of parities such that the useful life time period can be widened i.e., we have constant failure rates for longer period of time with coding. Another interesting observation is that coded system improves the wear-out period only if exceeding number of parities are used i.e. a rate of $1/10$ gives us a reasonable improvement although the aging can be greatly reduced at early life period for each component disk.

\begin{figure}[t!]
\centering
    \includegraphics[width=90mm, height=63mm]{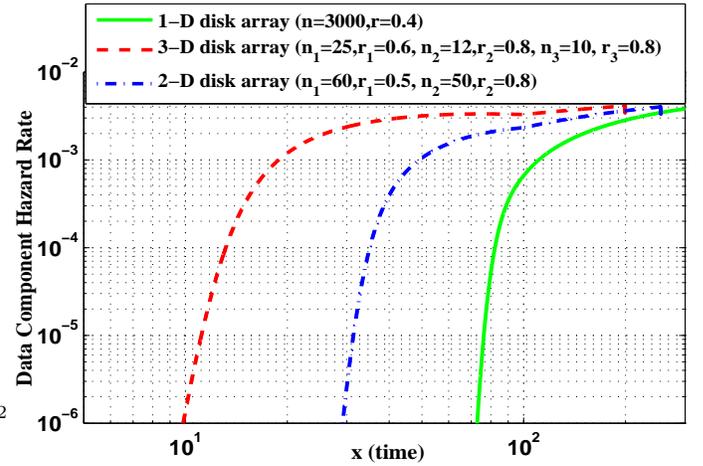}
    \captionof{figure}{\footnotesize Component hazard rate function using multidimensional MDS codes and structures. $\lambda(x)$ is assumed to be a simple bathtub curve obtained by using a composite distribution based on Weibull models. Number of disks in each system is 3000 and the total rate of each MDS code is 0.4.}
    \label{ComputeHazard3}
\end{figure}

In the next set of results, let us compare 1-D array of $n=300$ disks in which half of the disks  are dedicated to parity to a 2-D array of disks of the same size and relative redundancy, encoded with $(25,15,11)$ and $(12,10,3)$ MDS codes both in horizontal and vertical directions, respectively. Let us focus on array hazard rates i.e., data loss rate rather than individual disk hazard rates and assume independence. The results are presented in log-log scale in Fig. \ref{ComputeHazard2}. As can be seen, using a 2-D coding structure the data loss rate can greatly be lessened. In fact, since the block length and the rate of the component codes of the 2-D MDS code are reduced, some performance loss is observed at the infancy period. However, 2-D structure might be easier to implement as the MDS codes are of shorter length and larger rate are utilized.

Let us look at the individual data component disk hazard rates for a 3-D structure with an overall rate $0.4$ and $n=3000$ disks. Fig. \ref{ComputeHazard3} shows the data  component disk hazard rates for 1-D, 2-D and 3-D disk array structures of the same size.  Component MDS codes for 2-D structure have the parameters $(60,30,31)$ and $(50,40,11)$ whereas for 3-D structure, they have the parameters $(25,15,11)$, $(12,10,3)$ and $(10,8,3)$. As can be observed, although the array hazard rates show better performance with multi-dimensional MDS structures, the component hazard rates are worse compared to that of 1-D disk array. This is mainly due to 1-D MDS-protected system performs better (in fact it comes close to the performance lower bound) than  multiple short block length MDS codes of the same rate. However, the performance gain of multidimensional disk structures is rather in terms of low complexity due to using short block length and high rate MDS codes. In addition, if the array of disks fail for 1-D structure, the whole system fails and data is lost. On the other hand, multidimensional structures have many arrays of shorter length and it is low probability to lose all of them at once. It is not hard to see that data component disk hazard rates of multidimensional structures (2-D and 3-D arrays in our case) are close to that of 1-D array, particularly for useful and wear-out time periods. This means that the lower bounds can be achieved using multidimensional structures for a large fraction of time of a drive's lifespan. Finally, we note that our arguments are based on a conventional decoding algorithm used for product codes, more advanced algorithms might improve the over all decoding performance.

\section{Conclusion}

Generalized expressions are given for disk arrays that are protected by MDS-parities. A brief analysis of the interaction is also presented between redundancy and aging of MDS-parity based disk
array systems in a distributed storage scenario. The relationship between redundancy level and aging
is demonstrated using general formulations and accurate
distributions that is more reflective of the real life failure scenarios. Asymptotic results show that performance lower bounds are achievable with large scale storage networks as long as independence is assumed among the component failures.
Although neat compact form expressions may not exist for some of the
derivations, numerical results provide some intuition about the
behavior of such disk arrays under independent failure modes. Results are extended to include  multidimensional disk arrays to show that there might be practical ways to get close to predicted performance lower bounds for the component hazard rates.

\appendices
\section{Proof of Theorem 1}
Let us begin this section with the following lemma.

\textbf{Lemma 7:} \emph{The function $\psi_t(z,n, \lambda(x))$ satisfies
the following relationship for any integer $z$, satisfying $0\leq z \leq n$
\begin{eqnarray}
\frac{\psi_t(z,n, \lambda(x))}{\psi_{t-z}(n-z, \lambda(x))} =
\binom{n}{z} \left(1-R(x)\right)^z \label{Thm1}
 \end{eqnarray}}
\emph{where $\psi_{t}(n, \lambda(x)) \triangleq \psi_{t}(0, n,
 \lambda(x))$ is the cumulative distribution function of the binomial distribution.}

 \textbf{\emph{Proof}:} First note that for $z > i$, we have the convention $\binom{i}{z} = 0$. Therefore we  rewrite the expression
 for $\psi_t(z,n, \lambda(x))$,
\begin{align}
 =& \sum_{i=z}^t \binom{n}{i} \binom{i}{z}
(1-R(x))^i R(x)^{n-i} \\
=& \sum_{i=z}^t \binom{n-z}{i-z} \binom{n}{z} (1-R(x))^i R(x)^{n-i}
\label{BinomIdentity} \\
=& \binom{n}{z} \left(1-R(x)\right)^z \sum_{i=z}^{t}
\binom{n-z}{i-z} (1-R(x))^{i-z} R(x)^{n-i}   \\
=& \binom{n}{z} \left(1-R(x)\right)^z \sum_{j=0}^{t-z}
\binom{n-z}{j} (1-R(x))^{j} R(x)^{n-z-j} \label{COV} \\
=& \binom{n}{z} \left(1-R(x)\right)^z \psi_{t-z}(n-z, \lambda(x))
 \end{align}
from which the result follows. Note that we make the change of
variables $j=i-z$ in  Eqn. (\ref{COV}) and the Eqn.
(\ref{BinomIdentity}) follows from binomial coefficient identity
$\binom{n}{i} \binom{i}{z} = \binom{n-z}{i-z} \binom{n}{z}$. \hfill
$\blacksquare$

After establishing a useful lemma, let us give the proof of \emph{Theorem 1}. It is clear that  a horizontal block failure will occur only if $t+1$ or more disks
 fail in the horizontal block of size $n$ disks\footnote{Since the controller of disk array systems can identify which disks are failed, MDS codes are used to correct erasures.}. Due to independence
 assumption, the
 reliability of such a block is given by
 \begin{eqnarray}
 S(x) = \sum_{i=0}^t \binom{n}{i} (1-R(x))^i
R(x)^{(n-i)} = \psi_t(n,\lambda(x)).
 \end{eqnarray}

 Let us find the probability density function of the time between component failures. This  is given by
 \begin{eqnarray}
 f_X(x) &=& -S'(x) = - \frac{d S(x)}{d x} \\
 &=& \sum_{i=0}^t  \binom{n}{i} \bigg\{ \frac{i R'(x)}{1-R(x)} (1-R(x))^i
R(x)^{(n-i)}
 \nonumber \\
 && - \frac{(n-i) R'(x)}{R(x)} (1-R(x))^i
R(x)^{(n-i)}
 \bigg\} \\
 &=& - n \frac{R'(x)}{R(x)}  \psi_t(n, \lambda(x)) \\
&& + \frac{R'(x)}{R(x)} \frac{1}{1-R(x)} \psi_t(1, n, \lambda(x)) \nonumber \\
&=& \lambda(x) n \psi_t(n, \lambda(x)) - \frac{\lambda(x) \psi_t(1,
n, \lambda(x))}{1-R(x)}
 \end{eqnarray}
where we used the fact that $R'(x)/R(x) = -\lambda(x)$. Thus,
finally from Eqn. (\ref{ODE}), we compute the total hazard
rate for all $k$ data components (a series of $k$ data storage units) as
follows\footnote{We note that the horizontal system failure rate is always equal to the sum of the component failure rates, regardless of the distributions used to describe the components. In other words, let $\lambda_H(x)$ be the failure rate of a horizontal system that consists of $N$ components. If the component failure rates are characterized by the set of hazard rates $\{\lambda_i(x)\}_{i=1}^N$, it is not hard to show that  $\lambda_H(x) = \sum_{i=1}^N \lambda_i(x)$. This result is used implicitly throughout the paper to find the per component hazard rates of multidimensional coded storage systems.},
\begin{align}
k \mu_c(x,n,r) =& \frac{f_X(x)}{S(x)} =  \frac{f_X(x)}{\psi_t(
n, \lambda(x))} \nonumber \\
=& \lambda(x) \left(n - \frac{\psi_t(1, n, \lambda(x))/\psi_t(n,
\lambda(x))}{1-R(x)}\right) \label{Fineqn}
\end{align}
If we use the result of Lemma 7 with $z=1$, i.e.,
\begin{align}
\psi_t(1,n,\lambda(x)) = n(1-R(x)) \psi_{t-1}(n-1,\lambda(x))
\label{z_1}
\end{align}
and Eqn. (\ref{Fineqn}) then we  arrive at Eqn. (\ref{Thm1Core}).

As for the lower bound,  we observe the following relationship due to $t
\geq i$,
\begin{eqnarray}
t\psi_t(0,n,\lambda(x)) \geq \psi_t(1,n,\lambda(x))
\end{eqnarray}
Using above equation and Eqn. (\ref{z_1}), we can proceed as follows,
\begin{eqnarray}
\mu_c(x,n,r) &\geq& \frac{\lambda(x)}{r}\left(1 -
\frac{t}{n(1-R(x))}\right) \\
&=& \frac{\lambda(x)}{r}\left(\frac{r - R(x)}{1-R(x)}\right)
\end{eqnarray}
from which the lower bound follows. Notice that if $R(x) > r$, then
this lower bound takes on negative values. Therefore,
the maximum operator is introduced to make the lower bound non-negative.
 \hfill $\blacksquare$

 \section{Proof of Theorem 4}

Let $R(a) = 1/q$ for some $a > 0, q > 1$, we have
\begin{align}
\psi_{t-1}(n-1,\lambda(a)) &= \sum_{i=0}^{t-1}\binom{n-1}{i} \left(1-\frac{1}{q}\right)^i \left(\frac{1}{q}\right)^{n-1-i} \nonumber  \\
&=  q^{-n+1}\sum_{i=0}^{t-1}\binom{n-1}{i} \left(\frac{q-1}{q}\right)^i \left(\frac{1}{q}\right)^{-i} \nonumber \\
&= q^{-n+1}\sum_{i=0}^{t-1}\binom{n-1}{i} (q-1)^i
\end{align}
and similarly,
\begin{eqnarray}
\psi_{t}(n,\lambda(a))  &=& q^{-n}\sum_{i=0}^{t}\binom{n}{i} (q-1)^i
\end{eqnarray}

Using an asymptotic result from coding theory that in a $q$-ary $n$ dimensional linear space  $\mathbb{F}_q^n$, Hamming spheres (balls) of radius $t$  can be bounded  for large $n$ and $t/n = 1-r \leq  1 - 1/q$ i.e., $r \geq  1/q$ by
\begin{eqnarray}
q^{(h_q(1-r)-o(1))n} \leq \sum_{i=0}^{t}\binom{n}{i}(q-1)^i  \leq  q^{h_q(1-r)n} \label{Ineq3}
\end{eqnarray}
where $h_q(p)$ is the $q$-ary entropy function given by
\begin{align}
h_q(p) \triangleq p\log_q(q-1) +  p \log_q\left(\frac{1}{p}\right) + (1-p) \log_q\left(\frac{1}{1-p}\right)
\end{align}
and $o(1) \rightarrow 0$ as $n\rightarrow \infty$. In the context of coding theory, $q$ is usually an integer representing the size of the alphabet  over which the code is defined. Here in our case, it is not to hard to show that Eqn. (\ref{Ineq3}) is valid for any value of $q \in \mathbb{R}$ as long as  $qr \geq 1$.  Finally, we observe that
\begin{eqnarray}
&& q^{h_q(1-\frac{rn}{n-1})(n-1) -o(n) -h_q(1-r)n   + 1}   \nonumber \\
&& \ \ \ \ \ \ \ \ \ \ \leq  \frac{\psi_{t-1}(n-1,\lambda(a))}{\psi_{t}(n,\lambda(a))} \leq  \nonumber \\
&& \ \ \ \ \  \ \ \ \ \ \ \ \ \ \ q^{h_q(1-\frac{rn}{n-1})(n-1) - h_q(1-r)n +o(n) + 1}  \label{EqnThm51}
\end{eqnarray}
and $\exists \epsilon > 0$ such that the following assures the convergence for large $n$,
\begin{align}
& Pr\Biggl\{ \Biggl| \frac{\psi_{t-1}(n-1,\lambda(a))}{\psi_{t}(n,\lambda(a))} \nonumber\\
& \ \ \ \ \ \ \ \ \ \ \ - q^{(n-1)h_q(1-\frac{rn}{n-1}) - nh_q(1-r)+1} \Biggl| > \epsilon \Biggl\}  = 0
\end{align}

Let us use the definition for $h_q(p)$ to expand our expression along with the asymptotical results that $\lim_{n \rightarrow \infty} \log_q(1 - \frac{rn}{n-1} ) = \log_q(1 - r)$ and $\lim_{n \rightarrow \infty} \log_q( \frac{rn}{n-1} ) = \log_q(r)$
\begin{align}
& \log_q(q^{h_q(1-\frac{rn}{n-1})(n-1) - h_q(1-r)n + 1}  ) \nonumber \\
& \ \ \ \ \ \ \ \ \ \ = (n-1)h_q(1-\frac{rn}{n-1}) - nh_q(1-r)  + 1
\end{align}
where
\begin{align}
& (n-1)h_q(1-\frac{rn}{n-1})  \nonumber \\
&    = (n-1) \Biggl[ \left(1-\frac{rn}{n-1}\right) \log_q(q-1)  \notag\\  & \ \ \ \ \ \ \ - \left(1-\frac{rn}{n-1}\right) \log_q\left(1-r\right) - \left(\frac{rn}{n-1}\right) \log_q\left(r\right)\Biggl]  \\
& = (n-1-rn)\log_q(q-1) - (n-1-rn) \log_q\left(1-r\right) \notag \\ & \ \ \ \ \  -rn \log_q\left(r\right) \label{Log1}
\end{align}
and similarly,
\begin{align}
nh_q(1-r)  =& \left(n-rn\right) \log_q(q-1) - \left(n-rn\right) \log_q\left(1-r\right) \notag \\ &   - rn \log_q\left(r\right) \label{Log2}
\end{align}

Finally, let us use Eqns (\ref{Log1}) and (\ref{Log2}) to obtain,
\begin{align}
&(n-1)h_q(1-\frac{rn}{n-1}) - nh_q(1-r)  + 1 \notag \\
& \ \ \ \ \ \ \ \ \ \ \ = 1 - \log_q(q-1) + \log_q(1-r)  \\
& \ \ \ \ \ \ \ \ \ \ \ = \log_q \left(\frac{q(1-r)}{q-1}\right)
\end{align}

This result justifies that, for large enough $n$, we have the following convergence
\begin{align}
q^{(n-1)h_q(1-\frac{rn}{n-1}) - nh_q(1-r)+1} \rightarrow \frac{q(1-r)}{q-1}
\end{align}
which completes the proof for $r \geq 1/q$. For $r \leq 1/q$, we have
\begin{align}
\frac{q(1-r)}{q-1} \geq 1 \label{EqnFirst}
\end{align}

On the other hand, It is easy to see that for $t \leq n-1$
\begin{align}
\sum_{i=0}^t \binom{n-1}{i} (q-1)^{i+1} &\geq \sum_{i=0}^{t-1} \binom{n-1}{i} (q-1)^{i+1} \\
&= \sum_{i=0}^{t} \binom{n-1}{i-1} (q-1)^{i} \\
&= \sum_{i=0}^{t} \left[ \binom{n}{i} - \binom{n-1}{i} \right] (q-1)^{i}
\end{align}
from which we obtain,
\begin{align}
q \sum_{i=0}^{t} \binom{n-1}{i} (q-1)^{i} \geq  \sum_{i=0}^{t} \binom{n}{i} (q-1)^{i} \label{EqnAux}
\end{align}

Using a similar argument, we can show that
\begin{align}
\frac{\psi_{t-1}(n-1,\lambda(a))}{\psi_{t}(n,\lambda(a)) } = \frac{q}{q-1} \left(1 - \frac{\sum_{i=0}^{t} \binom{n-1}{i} (q-1)^{i}}{ \sum_{i=0}^{t} \binom{n}{i} (q-1)^{i} } \right)
\end{align}
and using Eqn. (\ref{EqnAux}), we obtain
\begin{align}
\frac{\psi_{t-1}(n-1,\lambda(a))}{\psi_{t}(n,\lambda(a)) } \leq 1 \label{EqnSecond}
\end{align}

Combining Eqn. (\ref{EqnFirst}) and Eqn. (\ref{EqnSecond}), we have $\frac{\psi_{t-1}(n-1,\lambda(a))}{\psi_{t}(n,\lambda(a)) } = 1$ for $n \rightarrow \infty$. Finally, this implies $\mu_c(a,n,r) \rightarrow 0$ as $n \rightarrow \infty$ if $r \leq 1/q$.  \hfill $\blacksquare$

 \section{Proof of Theorem 5}

Before proving Theorem 5, let us
first prove the following useful lemmas.

\textbf{Lemma 8:} \emph{The ratio of an incomplete Gamma function to a complete
Gamma function satisfies the following relationship,}
\begin{eqnarray}
1 - \frac{b}{a-b-1}< \frac{\Gamma(a,b)}{\Gamma(a)} =
\frac{\Gamma(a,b)}{(a-1)!} < 1
\end{eqnarray}
\emph{where $\Gamma(a,b) =
\int_{b}^{\infty} t^{a-1} e^{-t} dt$ is the incomplete Gamma
function.}

 \textbf{\emph{Proof}:} Let us explore the difference,
 \begin{eqnarray}
 \Gamma(a) - \Gamma(a,b) &=& \int_{0}^{b} t^{a-1}e^{-t} dt \nonumber \\
 &<&  \int_{0}^{b} b^{a-1}e^{-b} dt \label{InEq1}\\
 &=& b^a e^{-b} \nonumber \\
 &=& \frac{b}{a-b-1} \int_{b}^{a-1} b^{a-1} e^{-b}dt \nonumber \\
 &<& \frac{b}{a-b-1} \int_{b}^{a-1} t^{a-1} e^{-t}dt  \label{InEq2} \\
  &<& \frac{b}{a-b-1} \int_{0}^{\infty} t^{a-1} e^{-t}dt \\
  &=& \frac{b}{a-b-1}  \Gamma(a)
 \end{eqnarray}
 which establishes the lower bound. The upper bound follows from the definition of incomplete beta function. Note that the integrand $t^{a-1}e^{-t}$ achieves its maximum at $t=a-1$ and for $0 < t < a-1$, it is increasing. Thus, for $0 < t < b$, we have the inequality (\ref{InEq1}) and for $b < t < a-1$ we have the inequality (\ref{InEq2}).
 \hfill $\blacksquare$

 Lemma 8 indicates that for a fixed $b>0$ and $a \rightarrow \infty$,
 $\Gamma(a,b) \rightarrow \Gamma(a)$. In addition, for a fixed $b>0$ and $a \gg b$,
we have the following approximation
\begin{eqnarray}
\frac{\Gamma(a,b)}{\Gamma(a)} \approx \frac{a-2b-1}{a-b-1} \label{ApproxLemma2}
\end{eqnarray}

\textbf{Lemma 9:} \emph{As $n \rightarrow \infty$ and
$R(x)\rightarrow 0$ while satisfying $\lim_{\substack{ n \rightarrow \infty \\ x \rightarrow \infty}}
nR(x) < \infty$, we have the following convergence}
\begin{align}
\psi_t(n,\lambda(x)) &\longrightarrow
 1-\frac{\Gamma(k,nR(x))}{\Gamma(k)} \approx \frac{nR(x)}{k-nR(x)-1} \label{Lemma21}
\end{align}
\emph{and as $n \rightarrow \infty$ and $x\rightarrow 0$ while
satisfying $\lim_{\substack{ n \rightarrow \infty \\ x \rightarrow 0}} n(1-R(x)) <
\infty$, we have the following convergence}
\begin{align}
\psi_t(n,\lambda(x)) &\longrightarrow
\frac{\Gamma(t+1,n(1-R(x)))}{\Gamma(t+1)} \approx \frac{2R(x) - r -1}{R(x)-r}\label{Lemma22}
\end{align}
\emph{where $R(x) = e^{-\int_0^x\lambda(y)dy}$}.

 \textbf{\emph{Proof}:} First note the following relationship,
\begin{eqnarray}
\psi_t(n,\lambda(x)) &=& \sum_{i=0}^t \binom{n}{i} (1-R(x))^i
R(x)^{n-i} \label{Lemma2P1} \\
&=& \sum_{j=n-t}^n \binom{n}{j} R(x)^j (1-R(x))^{n-j}  \nonumber\\
&=&1- \sum_{j=0}^{n-t-1} \binom{n}{j} R(x)^j (1-R(x))^{n-j}
\nonumber
\end{eqnarray}

If $\lim_{\substack{ n \rightarrow \infty \\ x \rightarrow \infty}} nR(x) < \infty$, then
 the asymptotical convergence of binomial distribution to Poisson distribution yields
\begin{eqnarray}
\binom{n}{j} R(x)^j (1-R(x))^{n-j} \rightarrow \frac{n^jR(x)^j
e^{-nR(x)}}{j!}
\end{eqnarray}

Thus, for sufficiently large $n$,  we have
\begin{eqnarray}
\psi_t(n,\lambda(x)) &=& 1 - \sum_{j=0}^{n-t-1} \frac{n^jR(x)^j
e^{-nR(x)}}{j!} \\
&=& 1 - \frac{\Gamma(n-t,nR(x))}{(n-t-1)!} \\
&=& 1 - \frac{\Gamma(k,nR(x))}{\Gamma(k)}
\end{eqnarray}
and using the approximation (\ref{ApproxLemma2}) with $a=k$ and $b=nR(x)$, the Eqn. (\ref{Lemma21}) follows. Note that if $n \rightarrow \infty$, then
$k = nr \rightarrow \infty$. Similarly, using Eqn.
(\ref{Lemma2P1}) and same line of proof, we can obtain Eqn. (\ref{Lemma22})   \hfill $\blacksquare$

Next, we give the proof of \emph{Theorem 5}. First, consider $n\rightarrow
\infty$ and $R(x) \rightarrow 0$ while satisfying
$\lim_{\substack{ n \rightarrow \infty \\ x \rightarrow \infty}} nR(x) < \infty$. Using the results of
Lemma 8 and Lemma 9, we can approximate the limiting
ratio,
\begin{align}
1 - \frac{\psi_{t-1}(n-1, \lambda(x))}{\psi_t(n, \lambda(x))} &\rightarrow
1-\frac{(n-1)R(x)}{nR(x)} \notag \\
& \ \ \ \ \ \ \times \frac{nr - nR(x)-1}{nr - (n-1)R(x)-1} \nonumber \\
&= \frac{nr-1}{n(nr-(n-1)R(x)-1)} \\
&\approx \frac{1}{n}
\end{align}

Therefore, we have $k\mu_c(x,n,r) \rightarrow \frac{k \lambda(x)}{nr} =
\lambda(x)$
which establishes what is asserted.

Similarly, let us consider
$n\rightarrow \infty$ and $x \rightarrow 0$ while satisfying
$\lim_{\substack{ n \rightarrow \infty \\ x \rightarrow 0}} n(1-R(x)) < \infty$.  Let $b = (n-1)(1-R(x))$ and use
lemma 8, Lemma 9 and the approximation (\ref{ApproxLemma2}). Employing some algebraic manipulations, we can obtain
\begin{align}
n\left(1-\frac{\psi_{t-1}(n-1, \lambda(x))}{\psi_t(n, \lambda(x))}\right) &\approx n\left(1- \frac{1-b/(t-b-1)}{1-\frac{b+1-R(x)}{t-b+R(x)-1}}\right) \\
&= \frac{R(x)(t-1) - t + b + 1}{(t-b-1)(t - 2b + 2R(x) - 2)}
\end{align}

Let us divide both the numerator and denominator by $n^2$. Letting $n \rightarrow \infty$, we can approximate the limiting
ratio as follows
\begin{eqnarray}
n\left(1-\frac{\psi_{t-1}(n-1, \lambda(x))}{\psi_t(n, \lambda(x))}\right)
&\rightarrow& \frac{r(1-R(x))}{(R(x)-r)(2R(x)- r - 1)} \nonumber
\end{eqnarray}

Therefore using \emph{Theorem 1,} we have
\begin{eqnarray}
k\mu_c(x,n,r) &=& n\lambda(x)\left(1-\frac{\psi_{t-1}(n-1, \lambda(x))}{\psi_t(n, \lambda(x))}\right) \\
&\rightarrow& \frac{\lambda(x)r(1-R(x))}{(R(x)-r)(2R(x)- r - 1)}
\end{eqnarray}
which completes the proof. \hfill $\blacksquare$

\end{document}